\documentclass[aps,prc,11pt,reprint,twoside,tightenlines,nofootinbib]{revtex4-1}
\usepackage{graphicx}
\usepackage{amsmath}
\usepackage{amsfonts}
\usepackage{amssymb}
\usepackage[normalem]{ulem} 

\usepackage{color}

\newcommand{\be}{\begin{equation}}
\newcommand{\ee}{\end{equation}}
\newcommand{\ba}{\begin{eqnarray}}
\newcommand{\ea}{\end{eqnarray}}

\begin{document}

\title{The $K^- d \to \pi \Sigma n$ reaction revisited}

\author{Daisuke Jido} 
\affiliation{Yukawa Institute for Theoretical Physics, 
  Kyoto University, Kyoto 606-8502, Japan}
\affiliation{J-PARC Branch, KEK Theory Center,
Institute of Particle and Nuclear Studies, 
High Energy Accelerator Research Organization (KEK),
203-1, Shirakata, Tokai, Ibaraki, 319-1106, Japan}

\author{Eulogio Oset}
\affiliation{
Departamento de F\'{\i}sica Te\'orica and IFIC, Centro Mixto Universidad de Valencia-CSIC,
Institutos de Investigaci\'on de Paterna, Aptdo. 22085, 46071 Valencia,
Spain
}

\author{Takayasu Sekihara} 
\affiliation{Institute of Particle and Nuclear
  Studies, High Energy Accelerator Research Organization (KEK), 
  1-1, Oho, Ibaraki 305-0801,
  Japan}

\preprint{YITP-12-***, J-PARC-TH-012, KEK-TH-1559}
\pacs{14.20.Jn,25.80.Nv,13.75.Jz,12.39.Fe}
\keywords{Multiple scattering; Faddeev calculation; Watson formalism; Kaon induced production of $\Lambda(1405)$; Chiral unitary model}

\date{\today}

\begin{abstract} 
  The appearance of some papers dealing with the $K^- d \to \pi \Sigma
  n$ reaction, with some discrepancies in the results and a proposal to
  measure the reaction at forward $n$ angles at J-PARC justifies to
  retake the theoretical study with high precision to make accurate
  predictions for the experiment and extract from there the relevant
  physical information. We do this in the present paper showing
  results using the Watson approach and the truncated Faddeev
  approach. We argue that the Watson approach is more suitable to
  study the reaction because it takes into account the potential
  energy of the nucleons forming the deuteron, which is neglected in
  the truncated Faddeev approach. Predictions for the experiment are
  done as well as spectra with the integrated neutron angle.
\end{abstract}
\pacs{11.80.Gw, 12.38.Gc, 12.39.Fe, 13.75.Lb}

\maketitle

\section{Introduction}
\label{Intro}

The unexpected peak in the invariant mass of the $\pi \Sigma$ system
found in \cite{Braun:1977wd} in the $K^- d \to \pi \Sigma n$ reaction
at energies around $1420 \text{ MeV}$ was an interesting surprise
which gives support to the theory of two $\Lambda(1405)$ states
\cite{Jido:2003cb} as was discussed in \cite{Jido:2009jf}. In this
paper it was found that for kaons in flight the single scattering peak
and the double scattering were well separated, such that the double
scattering showed a clear peak due to the excitation of the
$\Lambda(1405)$. The process can only occur in nuclei. Indeed, the
single scattering $K^- p \to \pi \Sigma$ occurs for invariant masses
above the $K^- p$ threshold and does not show the resonance shape of
the $\Lambda(1405)$ since this one occurs below threshold. However, in
the deuteron, the initial kaon can collide with the neutron, give
energy to this neutron, hence losing energy such that in a rescattering
with the proton it can produce the $\Lambda(1405)$. Since the
production is done with a proton, one predicts that the
$\Lambda(1405)$ produced is the one that appears at energies around
$1420$ MeV and narrow in the theoretical framework of
\cite{Jido:2003cb}, which is also supported by all chiral dynamical
works on the issue (see \cite{Ikeda:2012au} for a recent update). The
work of \cite{Jido:2009jf} was extended in \cite{Jido:2010rx} to study
the $\Lambda(1405)$ production at the lower energies of DAFNE, where
the experiment was still predicted to be successful if forward neutrons
in coincidence were measured. Based on the experimental observation
and the calculations of \cite{Jido:2009jf, Jido:2010rx}, a proposal
has been done at J-PARC \cite{exp} looking for neutrons in the forward
direction.

In between, two more theoretical papers on the issue have appeared
with different results \cite{Miyagawa:2012xz, revai}. The first one uses a three
body approach for the final state, in the line of Faddeev equations
but truncated to second order. The authors claim that the peak
observed corresponds to a threshold effect. The second one considers full Faddeev
equations for the final three particles but only shapes and not absolute cross sections
are presented. In this latter case it is also suggested to divide the
cross section by a predicted background in order to visualize better
the signal of the resonance that in the calculation shows up clearly.
  
In the present work we present two studies based on the Watson
expansion and Faddeev equations, truncated to second order, and we
observe that the Watson expansion is more realistic than the truncated
Faddeev approach, since it considers the potential energy of the
nucleons in the deuteron, which is neglected in
\cite{Miyagawa:2012xz}. We show that the Watson expansion is
equivalent to the one used in \cite{Jido:2009jf}, thus reconfirming
the results of that work. In addition we make predictions for forward
neutrons, apart from the angle integrated cross sections, which should
be useful in the planning of the J-PARC experiment.

\section{Formulation}

%%%%%%%%%%%%%%
\begin{figure}
\centerline{\includegraphics[width=6.5cm,bb=0 0 184 90]{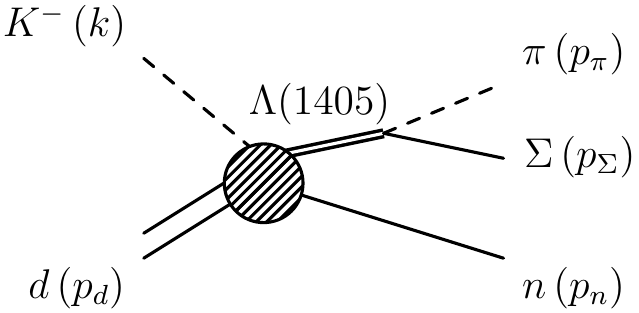}}
\caption{Kinematics of the $K^{-} d \to \pi \Sigma n$. \label{fig1}}
\end{figure}
%%%%%%%%%%%%%%

We consider the $K^{-}d \to \pi \Sigma n$ reaction, in which the
$\Lambda(1405)$ resonance is produced by the $\bar KN$ channel in the
intermediate state and decays to $\pi\Sigma$ being observed in the
final state (see Fig.~\ref{fig1}).  Because the $\Lambda(1405)$ is
located below the threshold of $\bar KN$, in order to create the
$\Lambda(1405)$ by the $\bar KN$ channel one needs nuclear
targets. Here we take a deuteron target, which is the simplest nucleus.
In this reaction, since the strangeness is brought into the system by
the incident kaon from the outside and the flow of the strangeness is
traceable, one can confirm that the $\Lambda(1405)$ resonance is
produced selectively by the $\bar KN$ channel.  This is an advantage of
the kaonic production over photo and pionic production in which the
strangeness should be created inside of the system and the
$\Lambda(1405)$ can be produced by both $\bar KN$ and $\pi \Sigma$
channels.  Here we consider in-flight incident kaons in order to avoid
the $\bar KN$ threshold contribution which contaminates the
$\Lambda(1405)$ spectrum (see Ref.~\cite{Jido:2010rx} for details).

%For the $\Lambda(1405)$ production with in-flight kaons, 

%%%%%%%%%%%%%%
\begin{figure}[t]
\begin{center}
\centerline{\includegraphics[width=8.5cm,bb=0 0 434 180]{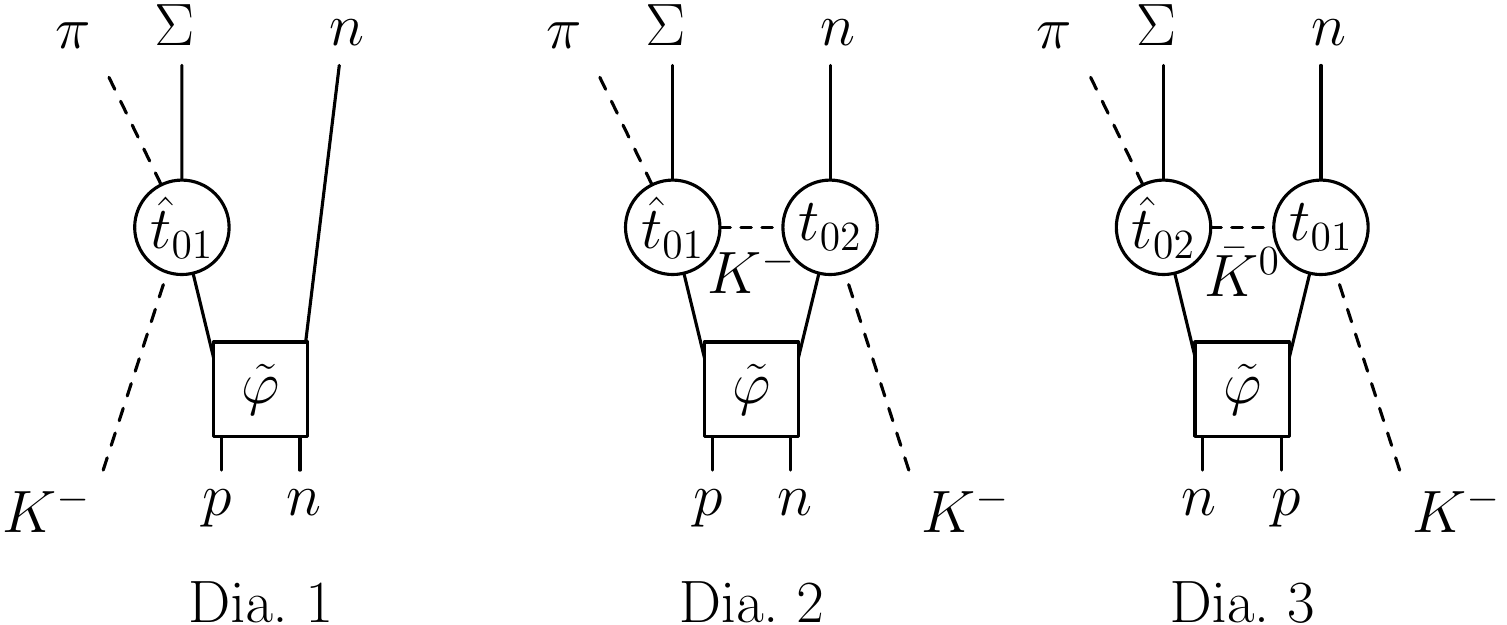}}
\caption{Diagrams for the calculation of the $K^{-}d \to \pi\Sigma n$ reaction.
\label{fig2}}
\end{center}
\end{figure}
%%%%%%%%%%%%%%

In Ref.~\cite{Jido:2009jf,Jido:2010rx}, the cross section of this process was calculated 
by considering the Feynman diagram shown in Fig.~\ref{fig2}. The total transition amplitude is given by summing up these three contributions:
\begin{equation}
    {\cal T} = {\cal T}_{1} + {\cal T}_{2} + {\cal T}_{3} .
\end{equation}
The left diagram 
of Fig.~\ref{fig2} corresponds to the impulse approximation in which the $\Lambda(1405)$ is produced by the incident kaon and the bound proton in the deuteron. The transition amplitude calculated in the rest frame of the deuteron target was obtained in Ref.~\cite{Jido:2009jf} as
\begin{equation}
   {\cal T}_{1} = \hat t_{01}(M_{\pi \Sigma}) \tilde \varphi \left(\vec p_{n} \right)
\end{equation}
with $\hat t_{01}$ the scattering amplitude of the $K^{-}p \to \pi
\Sigma$, $M_{\pi\Sigma}$ the invariant mass of $\pi\Sigma$ and $\tilde
\varphi$ the deuteron wave function in momentum space. The kinematical
variables are defined in Fig.~\ref{fig1}. The middle and right
diagrams shown in Fig.~\ref{fig2} are for the double scattering
contributions. There the kaon which scatters with one of the
nucleons of the deuteron rescatters with the other nucleon and creates
the $\Lambda(1405)$. In fact, it has turned out that these double
scattering processes dominate the $\Lambda(1405)$ production for
in-flight kaons because the energetic incident kaons can lose the
energy by kicking out one of the nucleons and create the
$\Lambda(1405)$ below the $\bar KN$ threshold. The transition
amplitudes were calculated in Ref.~\cite{Jido:2009jf} as
\begin{eqnarray}
   {\cal T}_{2} &=& \hat t_{01}(M_{\pi\Sigma}) \int \frac{d^{3}q}{(2\pi)^{3}} 
   \frac{\tilde \varphi(\vec q + \vec p_{n} - \vec k)}{q^{2} - m_{K}^{2} + i \epsilon}
   t_{02}(W_{1})  \label{t2} \ , \\
   {\cal T}_{3} &=& - \hat t_{02}(M_{\pi\Sigma}) \int \frac{d^{3}q}{(2\pi)^{3}} 
   \frac{\tilde \varphi(\vec q + \vec p_{n} - \vec k)}{q^{2} - m_{K}^{2} + i \epsilon}
   t_{01}(W_{1})   \label{t3}  \ ,
\end{eqnarray}
where $t_{01}$, $t_{02}$, $\hat{t}_{01}$, and $\hat t_{02}$ are the
two-body scattering amplitudes of $K^{-}p \to \bar K^{0} n$, $K^{-}n
\to K^{-}n$, $K^{-}p \to \pi \Sigma$, and $\bar K^{0} n \to \pi
\Sigma$, respectively, and the energies $q^{0}$ and $W_{1}$ are given
by
\begin{eqnarray}
   q^{0} &=& M_{N} + k^{0} - p^{0}_{n} \ , \label{q0} \\
   W_{1} &=& \sqrt{(q^{0}+p_{n}^{0})^{2} - (\vec q + \vec p_{n})^{2}} \ .
\label{W1}
\end{eqnarray}
The minus sign appearing in Eq.~(\ref{t3}) takes account of the
isospin configuration of the nucleons in the deuteron. The details of
the derivation of the transition amplitudes and the approximations
done were shown in Refs.~\cite{Jido:2009jf,Jido:2010rx}. In the
calculation given in Refs.~\cite{Jido:2009jf,Jido:2010rx} the two-body
scattering amplitudes are calculated by purely two-body dynamics based
on the chiral unitary approach, and they depend only on the invariant
mass carried by the interacting pair.

The issue raised in Ref.~\cite{Miyagawa:2012xz} is how one should
calculate the energy of the exchange kaon, $q^{0}$, in the loop of the
double scattering diagram with bound particles. Here it is not our
intention to derive an exact formulation which takes into account all
the contributions but to find an efficient approximation to treat the
bound nucleons by considering a few diagrams. One has the Faddeev
approach as one of the exact treatments of this reaction by
considering the $K^{-}d$ scattering as a three-body dynamical
problem. Therefore solving the equation following the Faddeev approach
to all orders with given two-body dynamics one would obtain an exact
solution of the $K^{-}d \to \pi \Sigma n$.

In fact, the prescription given by Eq.~(\ref{q0}) is based on the Watson formalism~\cite{Watson:1953zz} for reactions with bound particles. Here we compare the Watson and Faddeev approaches based on Ref.~\cite{Picklesimer:1983}. 
%The Watson approach for the three body problem is summarized in Ref.~\cite{Picklesimer:1983}. 
To make the formulation simpler let us consider a $K^{-} d \to \Lambda(1405) n$ transition in which the deuteron is a bound state of a proton and a neutron and the $\Lambda(1405)$ is a bound state of $\bar KN$. The coupled channels effect of $\bar KN$ and $\pi \Sigma$ is irrelevant to the present discussion and can be implemented into the two-body dynamics straightforwardly in certain ways. We assume that there exist only two-body forces between $p$, $n$ and $\bar K$, and we label $\bar K$, $p$ and $n$ in the initial state as particle 0, 1 and 2, respectively. The system is described by the total Hamiltonian of the three-body system
\begin{equation}
   H = K_{0} + K_{1} + K_{2} + v_{01} + v_{02} + v_{12} \ ,
\end{equation}
where $v_{ij}$ represents the potential between the particle $i$ and $j$ and $K_{i}$ is the kinetic energy operator for particle $i$. We take non-relativistic kinematics for simplicity and just the $K^{-}pn$ channel. Later on we shall generalize it to have $K^{-}p \to \pi \Sigma$ in the last step.  For the Watson formalism we define also the Hamiltonian of the unperturbed systems for the deuteron as
\begin{equation}
   H_{d} = K_{0} + ( K_{1} + K_{2} + v_{12}) = K_{0} + H_{12} \ .
\end{equation}

The transition operator for the $K^{-}d \to  K^{-} pn$ process can be written based on the Watson equation for the deuteron target with $A=2$ as
\begin{equation}
   T = T^{d}_{01} + T^{d}_{02} \label{Watson}
\end{equation}
with the coupled equations 
\begin{eqnarray}
   T^{d}_{01} &=& \tau_{01} + \tau_{01} G_{d} T^{d}_{02} \ ,  \label{d01} \\
   T^{d}_{02} &=& \tau_{02} + \tau_{02} G_{d} T^{d}_{01} \ .  \label{d02}
\end{eqnarray}
The operator $\tau_{0i}$ satisfies 
\begin{equation}
   \tau_{0i} = v_{0i} + v_{0i} G_{d} \tau_{0i} = v_{0i} + \tau_{0i} G_{d} v_{0i} 
   \label{tau}
\end{equation}
with the resolvent of $H_{d}$ 
\begin{equation}
    G_{d}\equiv [E-H_{d}+i\epsilon]^{-1} \ . 
\end{equation}
Note that the $\tau_{0i}$ is obtained with the unperturbed resolvent $G_{d}$ for the deuteron in the Watson approach instead of the free three-body Green's function in the Faddeev approach. 

The transition operator can be also written in the multiple scattering structure 
\begin{equation}
    T = \tau_{01} + \tau_{02} + \tau_{01} G_{d} \tau_{02} + \tau_{02} G_{d} \tau_{01} + \cdots  \label{Wexp}
\end{equation}
Taking the deuteron wave function and the $K^{-}$ plain wave in the initial state and the plain waves for the three particles in the final state, we find that the first term of Eq.~(\ref{Wexp}) corresponds to the amplitude ${\cal T}_{1}$ and the third and forth terms do to ${\cal T}_{2}$ and ${\cal T}_{3}$ after approximating $\tau_{0i}$ as the free two-body scattering operator $t_{0i}$ given by the two-body scattering equation
\begin{equation}
   t_{ij} = v_{ij} + v_{ij} G_{0} t_{ij} = v_{ij} + t_{ij} G_{0} v_{ij} \label{freetwo}
\end{equation}
for $i \ne j$
with the free Green's function $G_{0} = [E-H_{0} + i\epsilon]^{-1}$. Here it is important to note that  in the Watson formalism the double scattering process is calculated with the Green's function $G_{d} = [E-K_{0}-K_{1}-K_{2}-v_{12} +i\epsilon]^{-1}$ in which the potential energy for the nucleons also appears together with the kinetic energies.   
Generalizing the approach to have $\pi\Sigma$ in the final state, the ${\cal T}_{2}$ amplitude can be obtained in the Watson formulation as
\begin{eqnarray}
   {\cal T}_{2} &=&  \langle \pi \Sigma n| t_{01} G_{d} t_{02} | K^{-} d \rangle \\
  & = & \int \frac{d^{3} q}{(2\pi)^{3}} \langle \pi (p_{\pi}) \Sigma (p_{\Sigma}) |t_{01}|K^{-}(q) p(p_{1}) \rangle 
  \nonumber \\ && \times
  \langle  K^{-} p n  | G_{d} | K^{-} p n  \rangle
  \nonumber \\ && \times
   \, \langle K^{-}(q) n(p_{n}) |t_{02}|  K^{-}(k) n (p_{2}) \rangle \varphi (\vec p_{2})
\end{eqnarray}
where the matrix element of the Green's function operator can be calculated as
\begin{widetext}
\begin{eqnarray}
\lefteqn{    
\langle  K^{-} p n  | G_{d} | K^{-} p n  \rangle = 
} && \nonumber \\  &=&
    \langle  K^{-} p n  |\frac{1}{E_{\rm tot} - K_{0}  - K_{1} - K_{2} - V_{12} + i\epsilon }| K^{-} p n  \rangle \\ &=& 
    \langle  K^{-} p n  |\frac{1}{E_{\rm tot} - K_{0}  - (K_{1}+\frac{1}{2}V_{12}) - (K_{2} + \frac{1}{2} V_{12}) + i\epsilon }| K^{-} p n  \rangle \\ &=& 
    \frac{1}{M_{d} + k^{0} - \omega_{K} - (M_{N}+ \frac{\vec p_{1}^{\, 2}}{2 M_{N}} +\frac{1}{2} V_{NN} ) -p_{n}^{0} + i\epsilon}  \label{intermi} \\
    &\equiv & \frac{1}{q^{0} - \omega_{K} + i \epsilon }
\end{eqnarray}
\end{widetext}
where $\omega_{K} = \sqrt{m_{K}^{2} + \vec q^{\, 2}}$ and
\begin{equation}
   q^{0} = M_{d} + k^{0}  - \left(M_{N}+ \frac{\vec p_{1}^{\, 2}}{2 M_{N}}+\frac{1}{2} V_{NN}\right) -p_{n}^{0} .
\end{equation}
In Eq.~(\ref{intermi}) we have considered that $K_{2}+ \frac{1}{2}
V_{12}$ should be the total energy of the outgoing neutron.  Recalling
that the sum of averages of the kinetic energy and potential of the
nucleon in the bound state is given by minus a half of the binding
energy and neglecting the small deuteron binding energy, we find that
$q^{0}$ is given by Eq.~(\ref{q0}). Therefore the
prescription~(\ref{q0}) is based on the Watson formalism.

The equivalent transition operators to Eq.~(\ref{Watson}) can be obtained in terms of $t_{ij}$ given in Eq.~(\ref{freetwo}), instead of $\tau_{ij}$ calculated with $G_{d}$, in the following way~\cite{Picklesimer:1983}:
Let us define 
\begin{equation}
   \tilde T_{ij}^{d} \equiv v_{ij} G G^{-1}_{d}  \label{Tdij}
\end{equation}
for $i\ne j$ with the full Green's function $G = [E - H + i\epsilon]^{-1}$. The operator $\tilde T^{d}_{01}$ and $\tilde T^{d}_{02}$ satisfy the coupled equations (\ref{d01}) and (\ref{d02}) with the help of the resolvent identity
\begin{equation}
   G = G_{d} + G_{d} (v_{01} + v_{02}) G \ .
\end{equation}
Therefore we find $\tilde T_{0i}^{d} = T_{0i}^{d}$. Hereafter we use $T_{ij}^{d}$ instead of $\tilde T_{ij}^{d}$. The full propagator can be also written as
\begin{equation}
   G = G_{0} + G_{0} (v_{01}+v_{02}+v_{12}) G  \ .  \label{fullG}
\end{equation}
Inserting Eq.~(\ref{fullG}) into Eq.~(\ref{Tdij}),
we have
\begin{equation}
   T_{01}^{d} = v_{01} G_{0} G_{d}^{-1} + v_{01} G_{0} (T_{01}^{d}+T_{02}^{d}+T_{12}^{d}) \label{T01dan} .
\end{equation}
%Eliminating $v_{01}$ using Eq.~(\ref{freetwo}) we obtain
Multiplying $(1+t_{01}G_{0})$ to both sides of Eq.~(\ref{T01dan}) from the left and using Eq.~(\ref{freetwo}), we obtain 
\begin{eqnarray}
   T_{01}^{d} &=& t_{01} G_{0} G_{d}^{-1} + t_{01} G_{0} (T_{02}^{d}+T_{12}^{d}) . \label{F01}
\end{eqnarray}
Similarly we have 
\begin{eqnarray}
   T_{02}^{d} &=& t_{02} G_{0} G_{d}^{-1} + t_{02} G_{0} (T_{01}^{d}+T_{12}^{d}) ,\label{F02} \\
   T_{12}^{d} &=& t_{12} G_{0} G_{d}^{-1} + t_{12} G_{0} (T_{01}^{d}+T_{02}^{d}) . \label{F12}
\end{eqnarray}
Equations (\ref{F01}), (\ref{F02}) and (\ref{F12}) are one of the expressions of the Faddeev equation~\cite{Faddeev:1960su,Alt:1967fx}.
As seen in these equations, the three-body transition operator in the Faddeev approach can be written as the free three-body Green's function $G_{0}$ and the free two-body scattering operator $t_{ij}$ given in Eq.~(\ref{freetwo}). 

If one makes the multiple scattering expansion of the transition operator given in the Faddeev approach, one finds
\begin{equation}
   T = t_{01} + t_{02} + t_{01} G_{0} t_{02} + t_{02} G_{0} t_{01} + \cdots  \label{Fexp} \ .
\end{equation}
where we have used $G_{0}G_{d}^{-1} = 1 - G_{0}v_{12} $. 
If we compare the multiple scattering expansions obtained in the Watson and Faddeev approaches, (\ref{Wexp}) and (\ref{Fexp}), we find that each term in the expansions is different in these two formulations, and especially, the scattering process in the Faddeev approach is calculated with the free Green's function.

\begin{figure}
\begin{center}
\includegraphics[width=0.45\textwidth,bb=0 0 703 374]{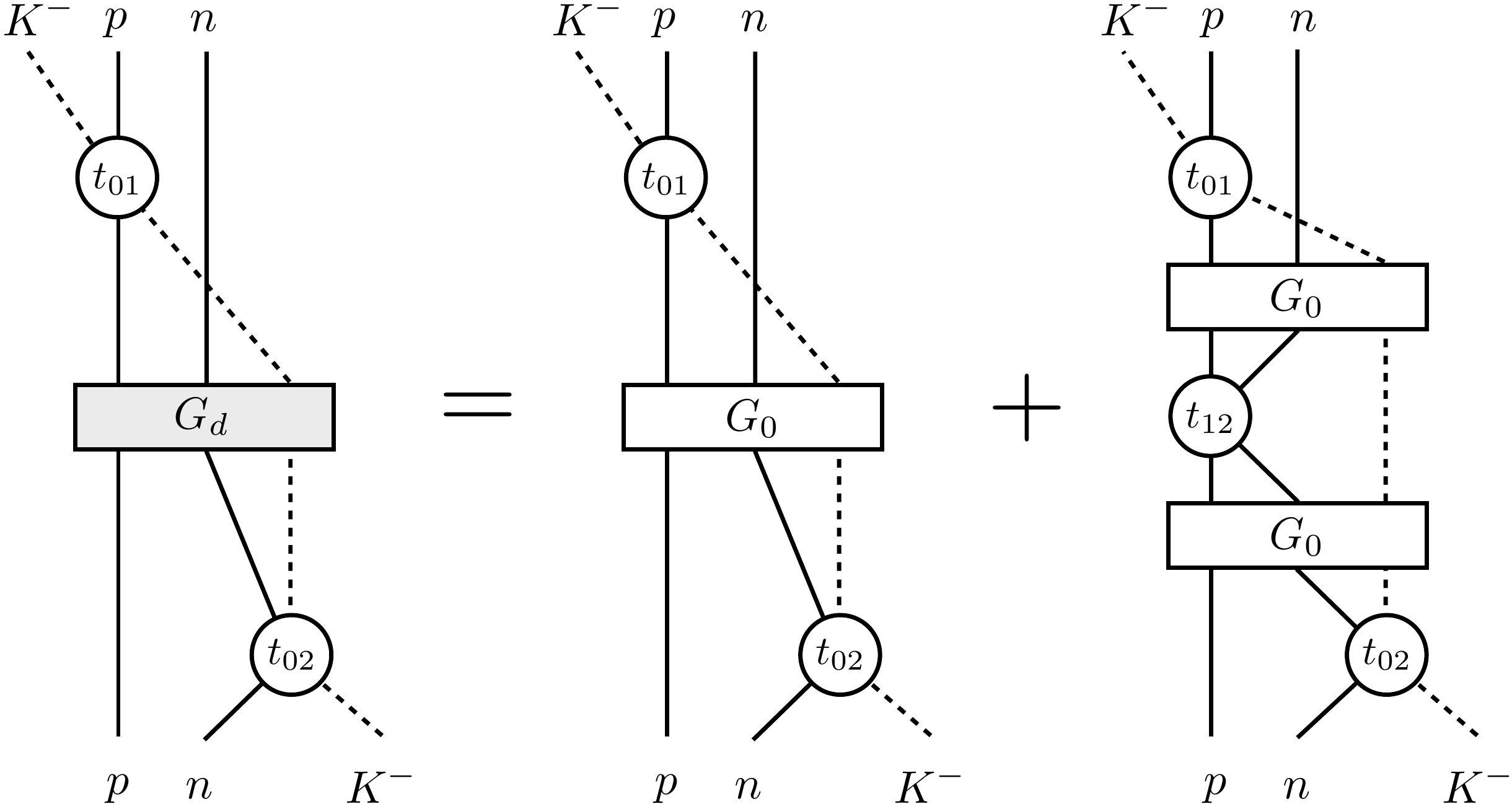} 
\caption{Schematic diagrams for Eq.~\eqref{eq:WatsonDouble}. The solid and dashed lines denote nucleons (proton and neutron) and kaon, respectively. $G_{d}$ and $G_{0}$ stand for the Green's function of the unperturbed deuteron system and the free three body, respectively, and $t_{ij}$ means the two-body transition amplitude.  The  double scattering contribution in the Faddeev approach contains only the middle diagram, while the double scattering in the Watson formulation have both middle and right diagrams.\label{fig:WatFad}}
\end{center}
\end{figure}

Let us write down the Watson amplitude \eqref{Wexp} in terms of $t_{ij}$ and $G_{0}$, which are the building blocks of the Faddeev approach. 
Considering Eq.~(\ref{freetwo}) and $G_{d} = G_{0} + G_{0} v_{12} G_{d}$, we have the identity 
\begin{equation}
   G_{d} = G_{0} + G_{0} t_{12} G_{0} \ . \label{Gd}
\end{equation}
Inserting this identity into Eq.~\eqref{tau} and recalling that $t_{0i} = v_{0i}/(1 - v_{0i}G_{0})$ we find 
\begin{eqnarray}
   \tau_{0i} &=& t_{0i} + t_{0i} G_{0} t_{12} G_{0} \tau_{0i} \label{tauFad} \\
   &=& t_{0i} + t_{0i} G_{0} t_{12} G_{0} t_{0i} 
   \nonumber \\ && \ \ \ 
   + t_{0i} G_{0} t_{12} G_{0} t_{0i} G_{0} t_{12} G_{0} t_{0i} + \cdots  
\end{eqnarray}
This implies that the single scattering term in the Watson formalism includes the multiple scattering terms in the Faddeev approach. Also for the double scattering term in the Watson approach, say $\tau_{02} G_{d} \tau_{01}$, using Eqs.~\eqref{Gd} and \eqref{tauFad}, we find that it includes more terms than the double scattering term in Faddeev approach.
%\begin{eqnarray}
%   \tau_{02} G_{d} \tau_{01} 
%  & = & t_{02} G_{d} t_{01} + t_{02}G_{0}t_{12}G_{0}t_{02}G_{d}t_{01} + \cdots \\
%  &=& t_{02} G_{0} t_{01} + t_{02} G_{0} t_{12} G_{0} t_{01} 
  %+ t_{02}G_{0} t_{12}G_{0}t_{02}G_{0} t_{01} 
%  + \cdots 
%\end{eqnarray}
Especially, in our approximation the double scattering terms are calculated with Eq.~(\ref{Gd}) as
\begin{equation}
   t_{01} G_{d} t_{02} = t_{01} G_{0} t_{02}+  t_{01} G_{0} t_{12} G_{0} t_{02} 
   \label{eq:WatsonDouble}
\end{equation}
where the first term corresponds to the double scattering term obtained in the Faddeev approach. Obviously, the Watson approach is better because it takes into account the interaction between the nucleons, as shown in Fig.~\ref{fig:WatFad}.
Finally we emphasize again that if one considers all of the terms of the multiple scattering expansion, namely if one solves the equations without truncation, one should get completely  identical solutions from both approaches. 

\section{Numerical results}

In the previous section we have discussed two approaches to describe
the $K^{-} d \to \pi \Sigma n$ reaction from the viewpoint of a
three-body dynamical problem.  One is the Watson approach and the
other is the Faddeev approach.  They are equivalent to each other if
one takes into account all orders of the multiple scatterings, but
they are different from each other if one truncates the multiple
scatterings at some finite order.  Especially the double scattering
process in the Watson approach has more terms than that in the Faddeev
approach.
Generally, each term of the multiple scattering expansion has different 
contributions in both approaches, but in practice it could give very similar
contributions in some systems with certain kinematical conditions.

In this section, we compare the double scattering terms obtained by 
the Faddeev and Watson approaches by taking the $K^{-} d \to \pi
\Sigma n$ reaction as an example.
In the present study we employ the chiral unitary
approach to evaluate the two-body meson-baryon dynamics and use
physical masses for ground-state hadrons, which slightly breaks the
isospin symmetry.  We consider scattering amplitudes for the $K^{-} d
\to \pi \Sigma n$ reaction up to the double scattering terms both in
the Watson and Faddeev approaches.  An important difference in the two
approaches is the treatment of the Green's function in the
intermediate states.  The single scattering term in our approximation
is exactly the same as that obtained in the Faddeev formulation.

In the Watson approach one uses the 
Green's function for the deuteron $G_{d}$ in the intermediate states,
therefore in the Watson approach the energy of the exchanged kaon in
the double scattering are expressed as,
\begin{equation}
q_{\rm (A)}^{0} = M_{N} + k^{0} - p_{n}^{0} , 
\label{eq:qA}
\end{equation}
to which we refer as case A.  This is the exchanged kaon energy in
double scattering which has been used in Refs.~\cite{Jido:2009jf,Jido:2010rx}.  
  
In contrast, in the Faddeev approach one uses the
free Green's function $G_{0}$ in the intermediate state, hence the
on-shell nucleons, which go for the second scattering, can appear in
the intermediate state in the double scattering.  In the case for the
double scattering, we have the expression for $q^{0}$ (case B) as,
\begin{equation}
q_{\rm (B)}^{0} = M_{N} + k^{0} - p_{n}^{0} 
- \frac{|\vec{q} + \vec{p}_{n} - \vec{k} |^{2}}{2 M_{N}} . 
\label{eq:qB}
\end{equation}
This is the exchanged kaon energy in double scattering which the
authors in Ref.~\cite{Miyagawa:2012xz} have used.  

The difference of two expressions for the exchanged kaon energy $q^{0}$ 
can be interpreted as how to implement the
binding effect on two nucleons in a deuteron.  Namely, in the
Watson approach the potential between proton and neutron is taken into
account and hence the secondary scattering nucleon in the intermediate
state keeps off-shell in the double scattering. Equation~\eqref{eq:qA} can 
be understood so that the kinetic energy of the secondary scattering nucleon 
is cancelled almost by the potential energy owing to the small binding energy.
In contrast, in the
Faddeev approach the effect of the bound nucleons is not considered and 
the secondary scattering nucleon in the intermediate state goes on-shell, 
meaning that the potential energy is neglected in the double scattering and
the binding of nucleons would be accounted separately in other terms of
the multiple scattering expansion.

%%%%%%%%%%%%%%
\begin{figure}[t]
\begin{center}
\centerline{\includegraphics[width=8.6cm,bb=0 0 432 302]{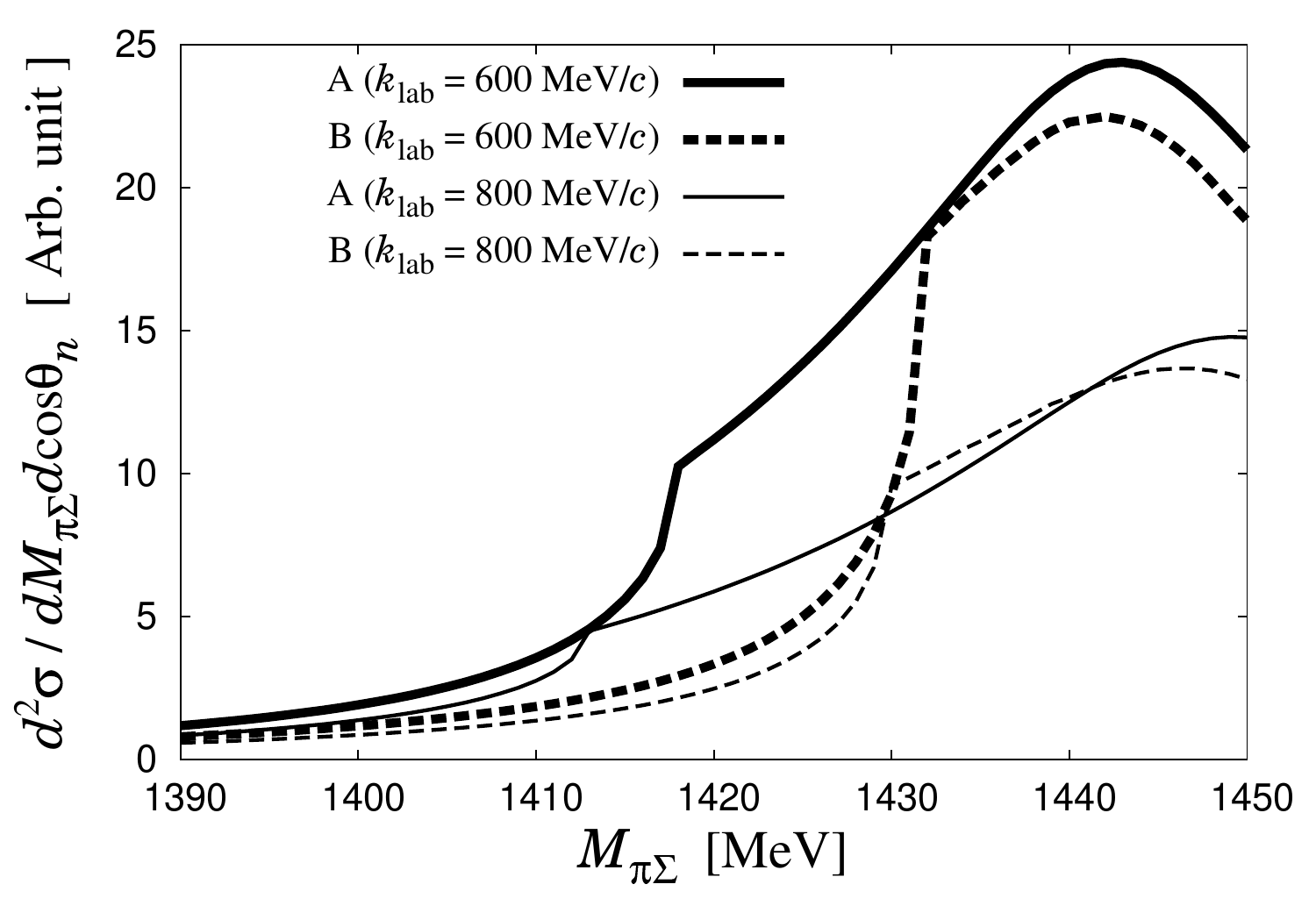}}
\caption{Differential cross section $d^{2} \sigma / d M_{\pi \Sigma} d
  \cos \theta _{\rm cm}^{n}$ coming from diagram 2 in Fig.~\ref{fig2} for 
  the reaction $K^{-} d \to \pi ^{-}
  \Sigma ^{+} n$ with a constant $\bar{K} N \to \pi \Sigma$
  amplitude~\eqref{eq:const_amp}.  Here we take two initial kaon
  momenta, $k_{\rm lab}=600$ and $800 \text{ MeV}/c$, and consider two
  intermediate kaon energy $q^{0}$ [A~\eqref{eq:qA} and
  B~\eqref{eq:qB}] coming from the Watson and Faddeev approaches,
  respectively.
  \label{fig3}}
\end{center}
\end{figure}
%%%%%%%%%%%%%%

Now let us perform numerical calculations of the single plus double
scattering for the $K^{-} d \to \pi \Sigma n$ reaction by using
$q^{0}$ of the above prescriptions A and B.  For simplicity we approximate
$W_{1}$ in Eq.~\eqref{W1} as,
\begin{equation}
W_{1} \approx \sqrt{(M_{N} + k^{0})^{2} - |\vec{k}|^{2}} , 
\end{equation}
bearing in mind that the first scattering amplitude, $t_{01}$ and
$t_{02}$ [see Fig.~\ref{fig2} and Eqs.~\eqref{t2} and \eqref{t3}], do
not make particular structures in the cross sections.  

First of all,
we consider the differential cross section for the center-of-mass
neutron scattering angle $\theta _{\rm cm}^{n} = 0^{\circ}$, on which
the authors in Ref.~\cite{Miyagawa:2012xz} concentrated.  Here, in
order to see the structure created by the underlying kinematical
features of the amplitudes rather than by the shape of the $\Lambda
(1405)$, we take the $\bar{K} N \to \pi \Sigma$ scattering amplitude
appearing in the second scatterings as,
\begin{equation}
\hat t_{01} = \hat t_{02} = \text{const.}
\label{eq:const_amp}
\end{equation}
The result of the differential cross section coming from diagram 2 of Fig.~\ref{fig2}
for the $K^{-} d \to \pi
^{-} \Sigma ^{+} n$ reaction with a constant $\bar{K} N \to \pi \Sigma$
amplitude is plotted in Fig.~\ref{fig3}, which corresponds to Fig.~9
of Ref.~\cite{Miyagawa:2012xz}.  The initial kaon momentum $k_{\rm
  lab}$ is fixed as $k_{\rm lab}=600$ and $800 \text{ MeV}/c$.  From
Fig.~\ref{fig3}, we find a cusp structure in both cases of A and B.
The cusp structure comes from the three-body unitarity cut for the
intermediate $K^{-}pn$ system as pointed out in
Ref.~\cite{Miyagawa:2012xz}.  It is important noting that the cusp position 
depends on the prescription of the intermediate kaon energy, $q_{(A)}$ or $q_{(B)}$.
While the cross section rapidly rises
around $M_{\pi \Sigma} \sim 1430 \text{ MeV}$ in case B with initial
kaon momentum $k_{\rm lab}=600 \text{ MeV}/c$, this effect becomes
moderate in case A with the same $k_{\rm lab}$.  Bearing in mind that
the Watson approach contains more terms compared to the Faddeev
approach up to the double scattering, we expect that, even if the
cusp from the three-body unitary cut appears in the cross
section, it will be more moderate than predicted in
Ref.~\cite{Miyagawa:2012xz}.  This moderation can be interpreted to be
caused by the fact that a more accurate energy sharing of two bound
nucleons in the deuteron is accomplished with the nonperturbative
Green's function for the deuteron.  

From the results of the initial
kaon momentum $k_{\rm lab}=800 \text{ MeV}/c$ with constant $\bar{K} N
\to \pi \Sigma$ amplitude, we see that more moderate cusp structures
are obtained for higher initial kaon momentum, and the position shifts
to lower energies for higher momentum.  We have also checked the angle
$\theta _{\rm cm}^{n}$ dependence for the cusps and have found that
the cusp position moves to lower energies, which will be important
when we integrate the angle to obtain the mass spectrum for the
$\Lambda (1405)$.

%%%%%%%%%%%%%%
\begin{figure}[t]
\begin{center}
\centerline{\includegraphics[width=8.6cm,bb=0 0 432 302]{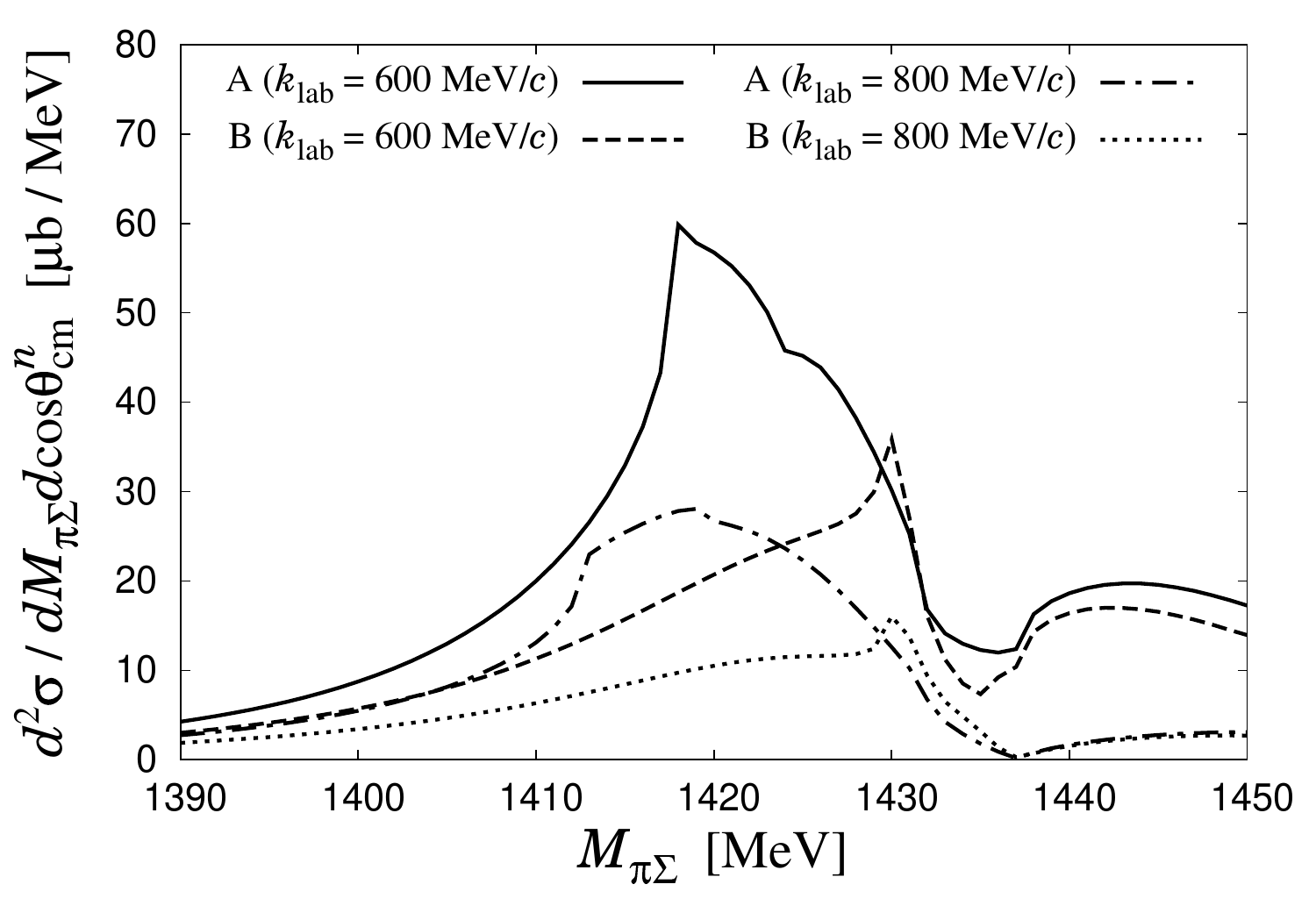}}
\caption{Differential cross section $d^{2} \sigma / d M_{\pi \Sigma} d
  \cos \theta _{\rm cm}^{n}$ of the reaction $K^{-} d \to \pi ^{-}
  \Sigma ^{+} n$ with $\bar{K} N \to \pi \Sigma$ amplitude in chiral
  dynamics.  Here we take two initial kaon momenta, $k_{\rm lab}=600$
  and $800 \text{ MeV}/c$, and consider two intermediate kaon energy
  $q^{0}$ [A~\eqref{eq:qA} and B~\eqref{eq:qB}] coming from the
  Watson and Faddeev approaches, respectively.
  \label{fig4}}
\end{center}
\end{figure}
%%%%%%%%%%%%%%

Next let us calculate the differential cross section $d^{2}\sigma / d
M_{\pi \Sigma} d \cos \theta _{\rm cm}^{n}$ at $\theta _{\rm cm}^{n} =
0 ^{\circ}$ with the actual $\bar{K} N \to \pi \Sigma$ amplitude for
$\hat{t}_{01}$ and $\hat{t}_{02}$ in chiral dynamics summing up all 
the three diagrams.  The results are
plotted Fig.~\ref{fig4} for the initial kaon momenta of $600 \text{MeV}/c$ 
and $800 \text{ MeV}/c$.  
The reason that we have two cusps in each line is that these come from 
the unitary cuts of the $\bar{K^{0}}nn$ and $K^{-}pn$ intermediate 
states of diagram 2 and 3, respectively. 
  In Fig.~\ref{fig4} we see that the
$\Lambda (1405)$ peak appears in case A in both momentum cases.
In spite of the cusps coming from the unitary cuts in
$d^{2}\sigma / d M_{\pi \Sigma} d \cos \theta _{\rm cm}^{n}$ at around
$M_{\pi \Sigma} \sim 1420 \text{ MeV}$, in case A this does not spoil
the $\Lambda (1405)$ spectrum.  The cusps in the differential cross
section become moderate in the higher momentum, as expected from
Fig.~\ref{fig3}.  In case B, on the other hand, the rapid rise plotted
in Fig.~\ref{fig3} around $1430 \text{ MeV}$ would spoil the $\Lambda
(1405)$ peak in the differential cross section.  However, as discussed
before, the rapid rise plotted in Fig.~\ref{fig3} and hence the
spoiled $\Lambda (1405)$ structure in Fig.~\ref{fig4} originate from
the use of the free Green's function in the intermediate state.  
Therefore, by taking more terms as the Watson approach does, 
one can obtain the $\Lambda (1405)$ peak in the differential
cross section.  Note, however, that even if the shape of the $\Lambda
(1405)$ is not well reproduced in case B, the presence of the
resonance has had an effect on the peaks, shifting the strength of
Fig.~\ref{fig3} from around $1430 \text{ MeV}$--$1450 \text{ MeV}$ to
$1420 \text{ MeV}$--$1430 \text{ MeV}$ in Fig.~\ref{fig4}.

%%%%%%%%%%%%%%
\begin{figure}[t]
\begin{center}
\centerline{\includegraphics[width=8.6cm,bb=0 0 432 302]{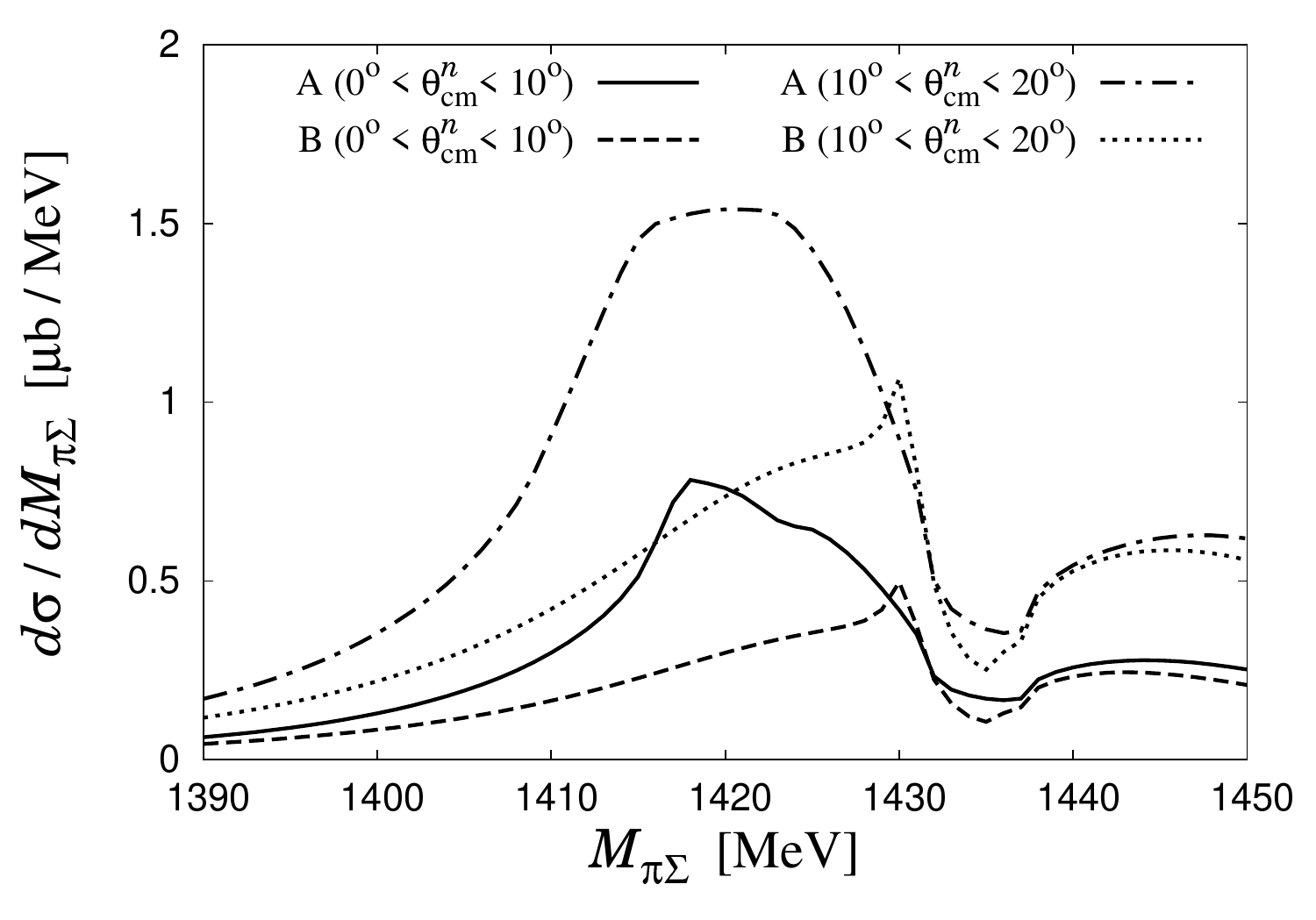}}
\caption{Invariant mass spectrum $d\sigma / d M_{\pi \Sigma}$ of the
  reaction $K^{-} d \to \pi ^{-} \Sigma ^{+} n$ with the initial kaon
  momentum $k_{\rm lab}=600 \text{ MeV}/c$.  Here the neutron
  scattering angle is integrated for two forward ranges, $0^{\circ} <
  \theta _{\rm cm}^{n} < 10^{\circ}$ and $10^{\circ} < \theta _{\rm
    cm}^{n} < 20^{\circ}$, and consider two intermediate kaon energies
  $q^{0}$ [A~\eqref{eq:qA} and B~\eqref{eq:qB}] coming from the
  Watson and Faddeev approaches, respectively.
  \label{fig5}}
\end{center}
\end{figure}
%%%%%%%%%%%%%%

Up to now we have considered only the limited forward angle of the emitted
neutron $\theta_{\rm cm}^{n}=0^{\circ}$. This is not realistic, because
in actual experiments one will observe the neutron in finite angles.  
Let us see the finite angle contribution. We 
calculate the invariant mass spectra by
integrating the angular dependence.  Since we
are especially interested in the forward scattering with small $\theta
_{\rm cm}^{n}$, where the $\Lambda (1405)$ production will be large due to
the double scattering processes~\cite{Jido:2009jf, Jido:2010rx}, we
plot in Fig.~\ref{fig5} the invariant mass spectrum at $k_{\rm
  lab}=600 \text{ MeV}/c$ for angles $0 ^{\circ} < \theta _{\rm
  cm}^{n} < 10 ^{\circ}$ and $10 ^{\circ} < \theta _{\rm cm}^{n} < 20
^{\circ}$.  From the figure, in case A, even in the scattering angle
with $0 ^{\circ} < \theta _{\rm cm}^{n} < 10 ^{\circ}$ the cusps at
the $\Lambda (1405)$ peak are smeared due to the angular dependence
for the cusps, and with $10 ^{\circ} < \theta _{\rm cm}^{n} < 20
^{\circ}$ the cusps are already negligible.  Since $\theta _{\rm
  cm}^{n} = 10 ^{\circ}$ in our condition corresponds to the neutron
scattering angle $\theta _{\rm lab}^{n} \sim 5^{\circ}$ in the
laboratory frame, this result indicates that the cusps structure does
not contaminate the $\Lambda (1405)$ peak structure if the neutron
detector in experiments is located in the forward angle of the
reaction with $\theta _{\rm lab}^{n} \gtrsim 5^{\circ}$.  For
comparison we also show the result for case B, which indicates that in
the Faddeev approach up to the double scattering the cusps depend
slightly on the scattering angle.

%%%%%%%%%%%%%%
\begin{figure}[t]
\begin{center}
\centerline{\includegraphics[width=8.6cm,bb=0 0 432 302]{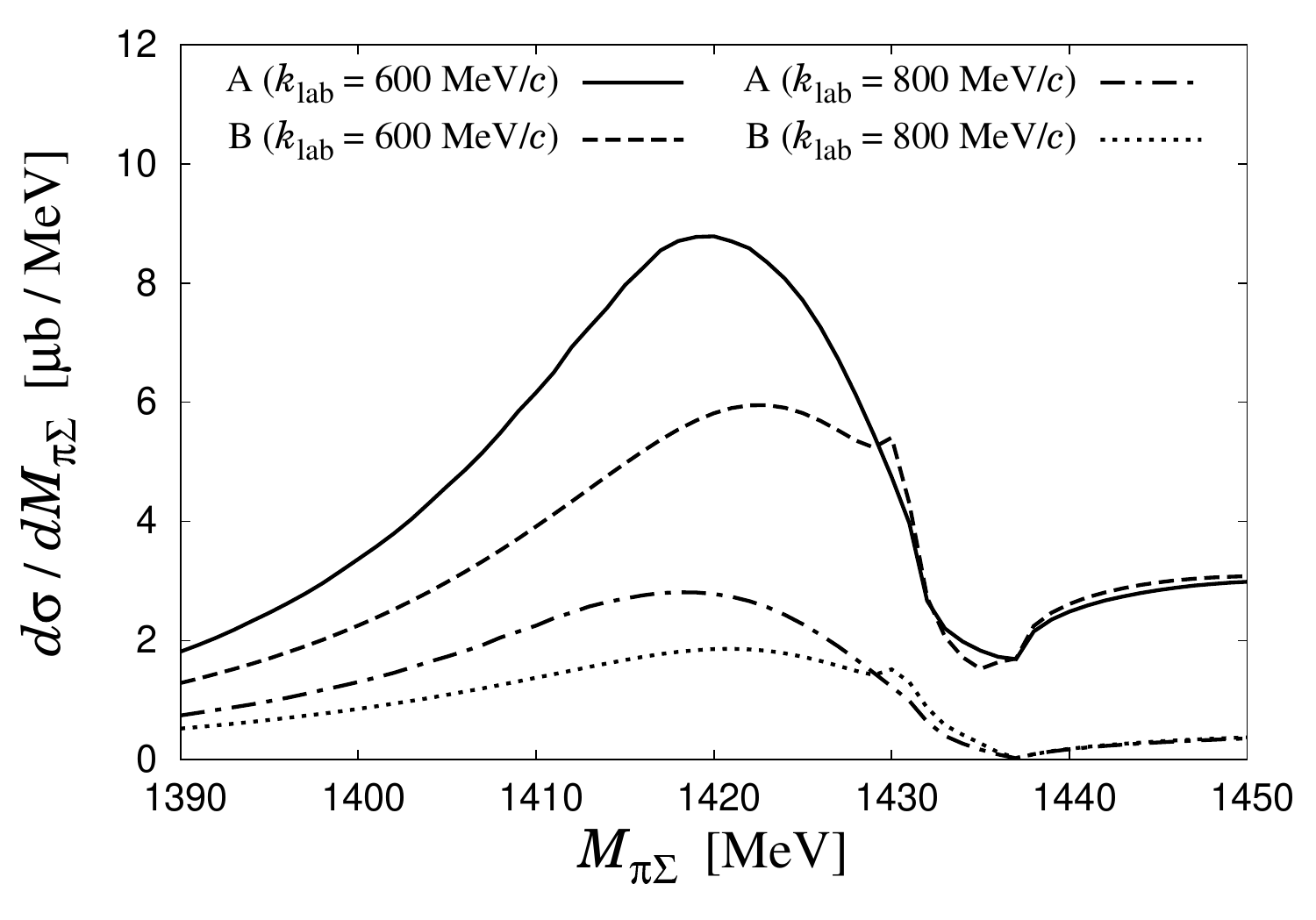}}
\caption{Invariant mass spectrum $d\sigma / d M_{\pi \Sigma}$ of the
  reaction $K^{-} d \to \pi ^{-} \Sigma ^{+} n$ integrated for the
  whole neutron scattering angle.  Here we take two initial kaon momenta,
  $k_{\rm lab}=600$ and $800 \text{ MeV}/c$, and consider two
  intermediate kaon energy $q^{0}$ [A~\eqref{eq:qA} and
  B~\eqref{eq:qB}] coming from the Watson and Faddeev approaches,
  respectively.
  \label{fig6}}
\end{center}
\end{figure}
%%%%%%%%%%%%%%

Finally we show the invariant mass spectrum integrated for the whole
angle in Fig.~\ref{fig6}.  In this case the cusp structures for case A having
appeared in previous figures completely disappear due to the angular
dependence for the three-body unitary cut, and we observe the $\Lambda (1405)$ peak
structure also in case B.

With respect to the paper of \cite{revai} a few comments are in order.
The input to construct the two body amplitudes is taken from
\cite{nina}, where an energy independent separable potential is
used. This neglects the important energy dependence in the interaction
provided by chiral theories. In any case, in the paper it is unclear
how the initial state $K^- d$ is produced and how does it couple to
the final state. It would be important to clarify this in view of our
arguments that the potential energy of the nucleons of the deuteron
plays a role in the energy denominators that appear in the final
formulas.  The paper adds a new interesting observation, realizing
that the kinematic peaks appear because of threshold of the reaction,
and then suggest to divide the cross section by another one that has
only background and no dynamical amplitudes. It is shown there that
after division by this new spectra the cross sections reflects the
properties of the resonance clearly.  There is only one problem to
it. This procedure is model dependent. Within a certain model the
kinematical spikes disappear with this procedure, but the spikes
depend on the model that one is using. If one wishes to apply the
method to a given experimental spectrum one is forced to choose some
model and the procedure is bound to create model dependent spikes
rather than eliminate them. Hence, the method is not advisable for an
experimental analysis.

\section{Conclusions} 

In this paper we have analyzed the $K^- d \to \pi \Sigma n$ reaction
for kaons in flight, looking for the cross section for forward neutron
angles for which a proposal is prepared for J-PARC. At the same time we
take advantage to introduce two new papers on the issue recently done
and discuss the meaning of their approach to the light of two
expansions, the Watson expansion of multiple scattering and the
truncated Faddeev approach. We realized that in the truncated Faddeev
approach the potential energy of the nucleons in the deuteron is
neglected. On the other hand, the Watson approach of multiple
scattering, which was used in \cite{Jido:2009jf}, takes into account this
information and represents a more suitable approach to the problem. 
Because in the Faddeev formalism the three particles are treated democratically,
it is insufficient to consider only the kaon double scattering contributions,
in which nucleon-nucleon interactions are not taken into account. Thus, 
when one considers a bound particle of two particles in the initial state 
in the Faddeev approach, one must consider contributions beyond the kaon double 
scattering contributions in order to take into account the $NN$ interaction
properly. In the Watson formulation, the bound particle is treated separately,
and thus one has an efficient multiple scattering expansion scheme. 
In any case we also showed that the peak observed in \cite{Miyagawa:2012xz}, related
there to threshold effects, is actually determined by the excitation
of the $\Lambda(1405)$ but somewhat distorted. This we could see by
changing the input and removing the $\Lambda(1405)$ in the amplitudes
that we use, and then we realize a substantial shift of the peak in
all the approaches. We also observe that in all approaches, when the
neutron angle is integrated, the threshold peaks are washed away and a
clear signal of the $\Lambda(1405)$ resonance shows up, though in the
Watson approach the curves are smoother and the predictions are deemed
more accurate.

  The present results clarify the situation and present a clear case for the experimental investigation of this reaction following the lines of the successful work of \cite{Braun:1977wd}.

\section*{Acknowledgments}

 This work is partly supported by the Grants-in-Aid for Scientific Research (No. 24105706,  and No. 22740161), DGICYT contract FIS2011-28853-C02-01, the Generalitat Valenciana in the program Prometeo, 2009/090, and the EU Integrated Infrastructure Initiative Hadron Physics 3 Project under Grant Agreement no. 283286. This work was done in part under the Yukawa International Program for Quark- hadron Sciences (YIPQS).

\end{document}